\documentclass{caosp306}

\usepackage{graphicx}
\usepackage{natbib}
\usepackage{threeparttable}
\articleNo{280}
\pubyear{2018}
\volume{48}
\volnumber{4}
\firstpage{1}
\received{May 9, 2018}
\accepted{June 22, 2018}

\begin{document}

\hauthor{M.A.\,Burlak, I.M.\,Volkov and N.P.\,Ikonnikova}

\title{Absolute parameters and period variation in a semidetached eclipsing variable V2247~Cyg}
%\subtitle{I. Overviewing the reasons, light curves solution}

\author{M.A.\,Burlak\inst{1}
          \and
          I.M.\,Volkov\inst{1,2}
          \and
          N.P.\,Ikonnikova\inst{1}
}

\institute{
              Sternberg Astronomical Institute, Lomonosov Moscow State University,
          Universitetskij Ave. 13, 119992 Moscow, Russia,
          \email{marina.burlak@gmail.com,ikonnikova@gmail.com}
          \and
              Institute of Astronomy of the Russian Academy of Sciences, 48 Pyatnitskaya street, 119017\, Moscow,\\
              Russia, \email{hwp@yandex.ru} }

\date{May 5, 2017}
\maketitle

\begin{abstract}
{We aim to determine the absolute parameters of the components of a
poorly studied Algol-type eclipsing binary V2247~Cyg. The data
analysis is based on our numerous precise $UBVRcIc$ photometric
observations and low resolution spectra. The photometric solution
reveals a semi-detached configuration with a less-massive
component filling its Roche lobe. The mass ratio, inclination,
effective temperatures, and fractional radii were determined. Masses
and radii of the components were obtained by a non-direct method.
The Eclipse Time Variation (ETV) diagram revealed the period
changes.}
\end{abstract}
%
%-------------------------------------------------------------------

\section{Introduction}
Algol-type stars (EA) are eclipsing binaries consisting of
spherical or ellipsoidal components and characterized by light
curves (LCs) in which it is possible to specify the beginning and
the end of eclipses. Algols are detached if both components are
inside corresponding Roche lobes and semi-detached if one of the
components fills its Roche lobe and loses matter. If the light
changes between eclipses due to the gravitational deformation, we
can estimate the mass ratio of the components without involving
spectral studies. The analysis of multicolour LCs of Algols (EA)
provides absolute parameters for their components which make
possible clarifying the evolutionary status of the eclipsing
binary.

The star BD~+33~4035 $V$=10.9 mag was designated as V2247~Cyg (EA)
in the 76th name-list of variable stars,
Kazarovetz et al.(2001), based on a private communication of
N.E.~Kurochkin and V.P.~Goranskij, 1999, hereinafter KG. Due to
the lack of data on the spectroscopic orbit of the object, we
estimated the absolute parameters such as semi-major axes, radii
and masses by the non-direct method described in Volkov et al. (2017). The precision of these values is not
better than 10 percents, but it is good enough to find out that the
location of the components on the diagrams given at the end of
this article coincides with that of other binaries with well known
parameters. Another aim of this work is to construct an ETV
diagram for the star using our own and archive observations and to
interpret the orbital period long-term change first found in this
study.

\section{Observations and data reduction}

{\sloppy \textbf{\textit{UBVRcIc} photometry}. The main set of $UBVRcIc$
observations of V2247~Cyg was obtained in 2013-2017 at the Crimean
station of the Sternberg Astronomical Institute (SAI) with the
0.5-m \textit{f}/4 Maksutov telescope (AZT-5) equipped with a CCD
camera Apogee Alta U16M. All reductions and aperture photometry
were made using the MaximDL software. The observational data were
reduced to the standard Johnson-Cousins photometric system according
to the following linear formulae:
\begin{equation}
\begin{array}{c}
U=U_{inst}+0.135((U_{inst}-B_{inst})-(U_{st}-B_{st}))-\\
-0.007((B_{inst}-V_{inst})-(B_{st}-V_{st})),\\
B=B_{inst}-0.039((B_{inst}-V_{inst})-(B_{st}-V_{st})),\\
V=V_{inst}+0.088((B_{inst}-V_{inst})-(B_{st}-V_{st})),\\
Rc=Rc_{inst}-0.043((B_{inst}-Rc_{inst})-(B_{st}-Rc_{st})),\\
Ic=Ic_{inst}-0.063((B_{inst}-Rc_{inst})-(B_{st}-Rc_{st})),\\
\end{array}
\end{equation}
where index $inst$ represents instrumental magnitudes of the
observed star and $st$ refers to the magnitudes of the standard
star. The reduction coefficients were obtained by averaging the
results of our observations of the standard area PG1633+099 of Landolt (1992) and observations of M67 performed by
D.Yu.~Tsvetkov.

}

The $UBVRIRc$ CCD photometer equipped with a VersArray~512UV
designed by one of the authors was used to make a calibration of the star
in the standard Johnson-Cousins system on August 11, 2017. An
equatorial standard GSC~543~227 was observed, whose $UBV$
magnitudes were taken from Landolt (2009) and $RI$
from Moffett \& Barnes (1979), where the star was designated as
113~466. The observations were carried out with a Zeiss-600
reflector located at Mt. Koshka, Crimean Astrophysical
Observatory(CrAO).

\textbf{Photoelectric observations.} The 0.6-m Zeiss
\textit{f}/12.5 Cassegrain telescope installed at the Crimean
station of SAI equipped with the $UBV$ photometer with
a photomultiplier EMI~9789 (PMT), constructed by Lyutyj (1971), was used to get full $UBV$ LCs. This
photometer has an instrumental system $U$ much closer to that of
Johnson than AZT-5 which has glass elements in its optical path
such as a meniscus and a corrector. So we used these observations to
check our AZT-5 $U$-observations and found good coincidence.

For both sets of observations with AZT-5 and PMT the only
reference star GSC~2695~1362($V_{st}$=11.24, $(B-V)_{st}$=0.40)
was used. No variability of the star was detected during the whole period of the observations.

The log of $UBVRcIc$ and $UBV$ observations is given in
Table~\ref{journal}.
%
%-------------Table1----------------
\begin{table}
\begin{center}
\caption{Log of photometric observations.} \label{journal}
\begin{tabular}{ccccc}
\hline\hline
Year& JD & N & System & Telescope\\
&2400000+...&&&\\
\hline
2013&56492-56495&1401&$V$&AZT-5, SAI\\
2014&56849-56857&2675&$UBVRcIc$&AZT-5, SAI\\
2015&57196-57240&5174&$UBVRcIc$&AZT-5, SAI\\
2017&57934-57980&2820&$UBVRcIc$&AZT-5, SAI\\
2017&57979-57994&25&$UBV$&PMT, SAI\\
\hline\hline
\end{tabular}
\end{center}
\end{table}
%--------------------------------------

\textbf{Spectral observations.} Low-resolution spectra of
V2247~Cyg were obtained at the 1.25-m reflector of the SAI Crimean
station. We used a diffraction spectrograph with a 600~lines/mm
grating. The slit width was 4$''$. The detector was an ST--402 CCD
($765\times510$ pixels of $9\times9 \mu$m). The spectral resolution
(FWHM) was 7.4~\AA. The spectra cover the wavelength interval from
3900 to 7200 \AA. V2247~Cyg was observed on July 21, 30, August
27, and October 12, 2017.

\textbf{Photographic observations.} One extra timing of the primary
minimum was obtained from an unpublished study of KG, who measured
the star's magnitudes using 165 photographic plates of the SAI
archive and derived a correct period of eclipses by the Lafler-Kinman method, Lafler \& Kinman (1965), $P=1^d.254861$.
These data one can find in the card catalogue of SAI.

The system is listed in the catalogues of eclipsing stars of
Malkov et al.(2006) and Avvakumova et al. (2013) which
contain no ephemeris for the object. Otero (2008) derived a period of $P=1^d.25486$ for the system, which perfectly
coincides with the KG value.

\section{Colour indices, spectra and determination of temperatures of the components}

\begin{figure}

\centerline{\includegraphics[width=\hsize,clip=]{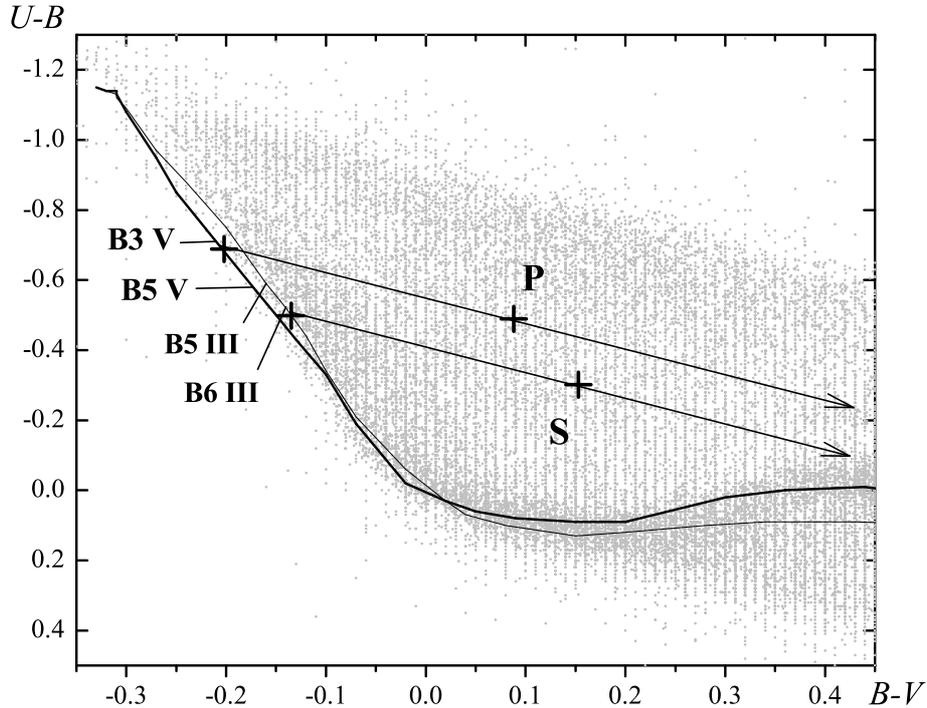}}
\caption{The $(U-B), (B-V)$ diagram. Arrows indicate the direction of
interstellar reddening. The bold line stands for the standard
luminosity class V sequence, and the thin one stands for that of
class III, Strai\v{z}ys (1992). Grey points represent
observations in the Johnson $UBV$ system from the
Mermilliod (1997) catalogue. Crosses mark observed and
dereddened indices of the primary (P) and secondary (S)
components.} \label{ubbv}

\end{figure}

The most important parameters for the current analysis are the
temperatures of the components that can be found in the following
way. The colour indices of the light loss in the primary and secondary
minima are calculated directly from the LCs in different passbands
with no additional assumptions. Observed colour indices of the
primary and secondary components are dereddened using a $(U - B),(B
- V)$ two-colour diagram, see Fig.~\ref{ubbv},
$E(U-B)/E(B-V)~=~0.710$ was accepted for the B5 spectral class from
Table~11 in Strai\v{z}ys (1992). The colour indices
calculated this way are applied to determine the temperatures of
the components with the help of well-known calibrations.
\begin{equation}
\begin{array}{c}
\mathrm{Primary:~} (U-B)_0=-0.689\pm 0.015 (B-V)_0=-0.202\pm 0.010,\\
E(B-V)=0.290\pm 0.012.\\
\mathrm{Secondary:~}
(U-B)_0=-0.499\pm 0.020 (B-V)_0=-0.135\pm 0.015,\\ E(B-V)=0.288\pm 0.014.\\
\end{array}
\end{equation}
For completeness we calculated mean colour indices in both minima
and maxima of the LC. They are presented in Table~\ref{colours}.
Keep in mind that these data are not dereddened.
%-------------Table2----------------
\begin{table}
\begin{center}
\caption{The observed colours in both minima and maxima.}
\label{colours}
\begin{tabular}{ccccc}
\hline\hline
Phase& $U-B$ & $B-V$ & $V-R_c$ & $R_c-I_c$ \\
\hline
0.0 &   -0.33&  0.133&  0.097&  0.144\\
0.25&   -0.43&  0.102&  0.072&  0.115\\
0.5 &   -0.45&  0.095&  0.064&  0.109\\
0.75&   -0.42&  0.111&  0.069&  0.116\\
\hline\hline
\end{tabular}
\end{center}
\end{table}

We derived $T_1=17100$~K and $T_2=13000$~K from
Flower (1996). Popper (1980) gives the
temperatures that are 2000~K higher for the primary and 800~K
higher for the secondary. We can see that the temperatures derived
from the calibrations may have an uncertainty up to 1000~K for
this range of temperatures. Equal values of interstellar reddening
for both components can be obtained only if we attribute the third
or fourth class of luminosity to the secondary component. We
compared the results with the data from available surveys. A new
review of interstellar extinction made by
Green et al. (2015) gives $E(B-V)=0.18\pm0.02$ for the
distance to the star $d=1960$~pc (see Table~\ref{t3}). From
Schlafly \& Finkbeiner (2011) and Schlegel et al. (1998) one
can obtain $E(B-V)=0.29\pm0.02$ for the total extinction along the
given line of sight in the Galaxy. Taking into account the
distance to the star  we get $E(B-V)=0.23\pm0.02$ from an equation
in Bonifacio (2000). The errors of these surveys seem
to be understated. Nevertheless, we argue that such a discrepancy
is not critical and we accept the value $E(B-V)=0.29\pm0.01$ that
follows from the photometry.

%-------------------Figure 1 --------------------------
%----------------------------------------------------------

\begin{figure}

\centerline{\includegraphics[width=\hsize,clip=]{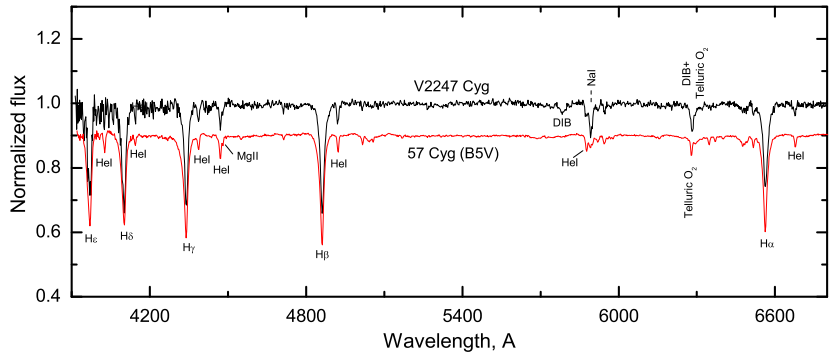}}
\caption{The spectra for V2247~Cyg and 57~Cyg normalized to the
continuum, obtained on July 30, 2017. The spectrum for 57~Cyg is
intentionally shifted along the vertical axis.} \label{sp}
\end{figure}

We obtained low-resolution spectra for the star in different
phases of the orbital period. We did not detect any significant
difference between the spectra obtained in the maximum and at the
bottom of the primary minimum. The spectrum is dominated by the
Balmer lines of hydrogen and lines of neutral helium. No lines of
ionized helium are seen. The presence of He\,I lines in the
absence of He\,II lines indicates the B spectral class
(Gray \& Corbally, 2009). The relative intensities of H\,I and
He\,I lines alone do not allow the precise temperature
determination due to possible helium abundance anomalies, which are
not rare in B-class stars. The
$\textrm{He\,I}\,\lambda4471/\textrm{Mg\,II}\,\lambda4481$ ratio
may be of use for specifying the temperature of stars later than
B3 and in our study it appeared very helpful as the lines of
Si\,III and Si\,IV, which are weaker, could not be resolved. Besides,
low spectral resolution and a low S/N ratio of our spectral data prevent a
precise luminosity classification. The comparison with the stars
of spectral classes between B3 and A0, and of luminosity classes
from V to III, yields the best agreement for the spectral class
B5\,V. The uncertainty is about one subclass. The spectrum of
V2247~Cyg is similar to that of 57~Cyg (B5\,V) (Fig.~\ref{sp}). A
B5\,V-III spectrum corresponds to a temperature of 15~400~K
according to the calibration of Strai\v{z}ys (1982), in
good coincidence with photometric results.

In addition to H\,I and He\,I absorptions in the spectrum of
V2247~Cyg, we identified prominent diffuse interstellar bands
(DIBs) centered at $\lambda$5780 and $\lambda$6284 and the strong
interstellar NaI D doublet. The strengths of these lines in
stellar spectra show a positive correlation with the observed
extinction, though there is a significant dispersion about the mean
relationship (Friedman et al., 2011; Herbig, 1993). To estimate interstellar extinction,
one needs to resolve Na\,I D1 and D2 lines, to measure their
strengths separately, and to control the D2/D1 ratio, because the
uncertainty is large at the higher Na column densities due to the
line saturation. The low resolution of our spectra does not allow
us to perform such a procedure. Similar problems arise if we
intend to estimate extinction from the equivalent widths of DIBs:
the DIB at $\lambda$6284 is blended with a telluric O$_2$ band
consisting of several absorption lines which are not detectable
with given resolution, and the DIB at $\lambda$5780 is too broad
and shallow to be measured securely, given a low S/N ratio.

%-------------------Figure 2 --------------------------

%----------------------------------------------------------

\section{Light curves solution}

\begin{figure}
\centerline{\includegraphics[width=\hsize,clip=]{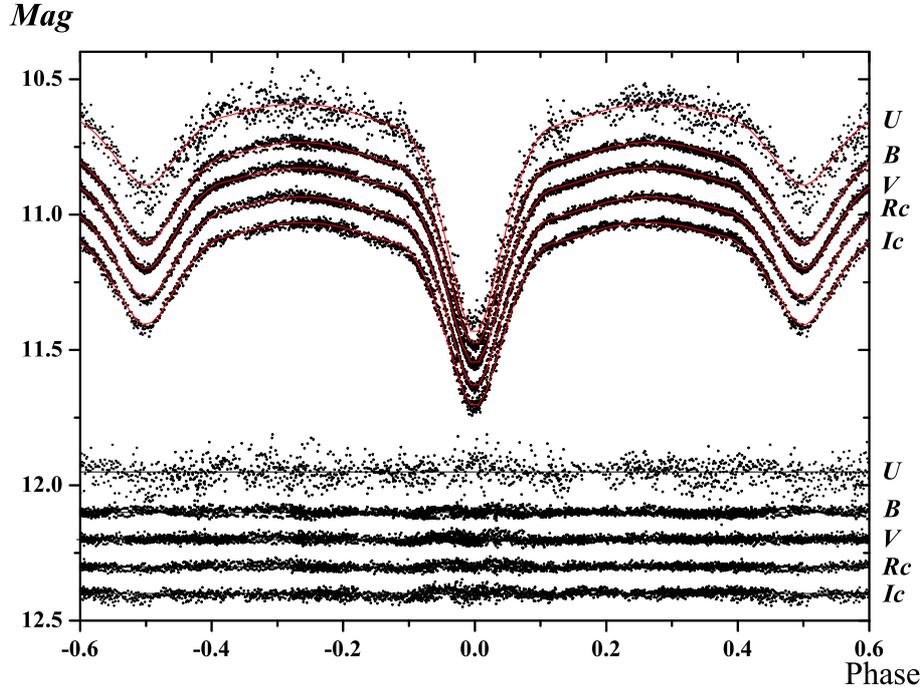}}
\caption{$UBVRcIc$ observations for V2247~Cyg. Red lines stand for
the best fits. Residuals from the best fits are shown at the
bottom of the picture.} \label{curves}
\end{figure}

The LCs of the binary show clear evidence for proximity effects,
see Fig.~\ref{curves}. So, we used the PHOEBE program of
Pr\v{s}a \& Zwitter (2005) a to analyze them. The best fit yielded
a semidetached system with the secondary filling its Roche lobe.
This agrees with the consideration inferred from our photometry
that the secondary component has moved far enough in its evolution
from the initial main sequence. We got an individual solution for
each passband of AZT-5 observations and then repeated the
procedure for less numerous PMT observations in 2017. All
solutions led to similar parameters: the mass ratio $q$, inclination
$i$, effective temperatures $T$, fractional radii $r_1, r_2$, and the potential $\Omega$. Mean weighted values of the parameters are
presented in Table~\ref{t3}. The solution of the LCs did not show
any presence of a third light and it was assumed to be zero,
$L_{3}=0$.

The temperatures of the components were included in solving LCs,
as the values obtained from calibrations seemed to have
discrepancies of up to 1000K for the considered temperature range,
see the previous section. We accepted a temperature of $T_1=17100$~K
as the first approximation for the primary component. Solutions
provided a confident minimum for the temperatures of the
components, $T_1=16500$~K and $T_2=11500$~K, and these are 500 and
1500~K lower than it follows from photometric calibrations. The found
temperatures correspond to the spectral types B4-5\,V and
B8\,III-IV in the calibration of Strai\v{z}ys (1982).

Taking into account high accuracy and a large amount of
observational data we included the albedo ($A$) of the secondary
component and the gravity brightening coefficients ($g$) as varied
parameters into the process of minimizing residuals. We found a
shallow minimum for $g_1$=$g_2$=0.90 and $A_2$=0.8, but its
reality is uncertain. So we assumed the theoretical values, see
Table~\ref{t3}. The albedo of the primary component has a
negligible effect on the shape of the LCs and was accepted from the PHOEBE of Pr\v{s}a \& Zwitter (2005). The limb-darkening
coefficients can be found from temperatures and gravitational
accelerations of the components. The best results were obtained
with a linear limb-darkening law. The coefficients were derived
with the PHOEBE program according to van Hamme (1993).
The solution is given in Table~\ref{t3}.
%----------------------Table 3----------------
\begin{table}
\begin{center}
\caption{Parameters derived from fitting the $UBVRcIc$ LCs.}
\label{t3}
\begin{tabular}{c c c}     % 2 columns
\hline \hline
Parameter         &  Primary         & Secondary           \\
\hline
$i$ [$^{\circ}$]  & \multicolumn{2}{c}{$79.08\pm 0.04$}  \\
$q(M_2/M_1$)      & \multicolumn{2}{c}{$0.812\pm 0.006$} \\
$T$ (K)           &  $16500\pm300$   & $11450\pm100$       \\
$Sp$              &  B4\,V           &  B7\,III-IV         \\
$BC$ (mag)        &  $-1.461$        & $-0.564$            \\
$\Omega$          &  $4.26\pm 0.02$  & $2.51$              \\
$A$               &   0.6            & 0.6                 \\
$g$               &   1.0            & 1.0                 \\
 \hline\hline
\end{tabular}
\end{center}
\end{table}
%-------------------------------------------------------

%-------------------Figure 3 --------------------------
%----------------------------------------------------------

Assuming a normal distribution for the residuals we get a mean
error for an individual observational point in every spectral
band:
\begin{center} $U-0.042$~(1305 points), $B-0.0107$~(2358 points), $V-0.0099$~ (3720 points),
$Rc-0.0100$~(2308 points), $Ic-0.0133$~(2326 points).
\end{center}

\begin{figure}
\centerline{\includegraphics[width=\hsize,clip=]{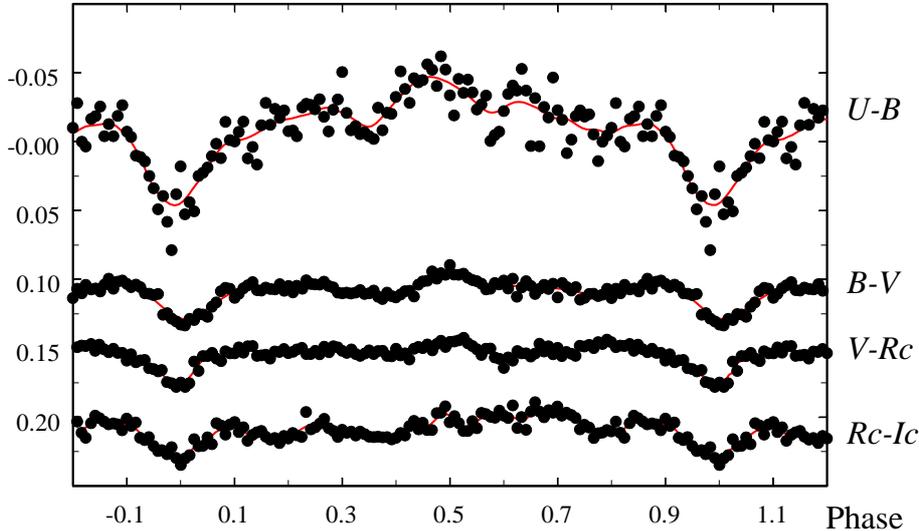}}
\caption{A plot showing colour indices (C.I.) phased with the
current period. The Y-axis labels correspond to $B-V$. Other C.I.
are shifted downwards, $U-B$ by 0.40, $V-Rc$ by 0.08 and $Rc-Ic$
by 0.09. Red lines are drawn by the Weighted Least Squares(WLS)
method.} \label{ubbvvrcrcic}
\end{figure}

Fig.~\ref{ubbvvrcrcic} contains the phased colour curves for the star.
The period was split into 120 intervals and the colour data were
averaged within each interval. This plot is of interest as it
demonstrates some asymmetry, especially in the $U-B$ colour index, which could be attributed to physical processes in the system.

\begin{figure}
\centerline{\includegraphics[width=\hsize,clip=]{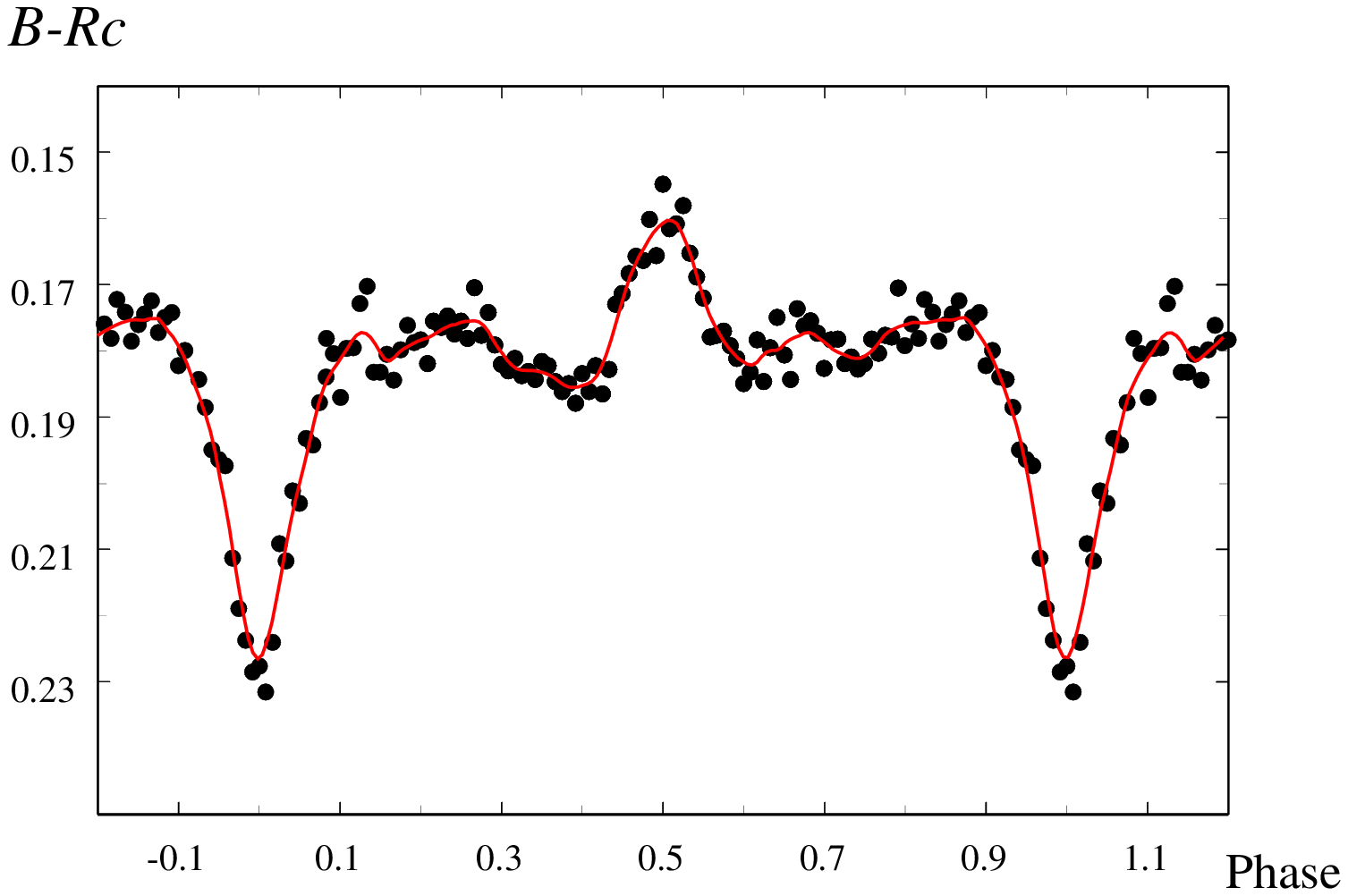}}
\caption{The $B-Rc$ colour index phased with the current period.
The red line is drawn by the WLS method.} \label{BRc}
\end{figure}

Fig.~\ref{BRc} shows the $B-Rc$ phased colour curve for V2247~Cyg which best describes the tendency of the binary to get redder when approaching the secondary minimum. If there were the reflection effect, the object would get bluer.

%-------------------Figure 4 --------------------------

%------------------------------------------------------
%-------------------Figure 5 --------------------------

%------------------------------------------------------

\section{Absolute parameters and ETV diagrams}

We found the masses of the components by a non-direct method we used in Volkov et al. (2017). The method is based on
the empirical mass-luminosity relation, the 3rd Kepler law, and
the relation between the absolute and relative radii:

\begin{equation}
\begin{array}{c}

\log(L_1/L_\odot)=3.99\log(M_1/M_\odot),\\
a^3/P^2=M_1(1+q),\\
R_1=r_1a.

\end{array}
\end{equation}

While the light curves solution provides a reliable $q$ value for
semidetached systems, see Table~\ref{t3}, the non-direct method
yields $q=0.81\pm0.01$, which is in perfect agreement with the
value from the LCs solution. The absolute parameters of the system are
presented in Table~\ref{t4}. Our photometric parallax,
$\pi=0.49\pm0.03$ mas, matches quite well the GAIA DR2 value,
$\pi=0.4310\pm0.0287$ mas, of Luri et al. (2018). We hope that with the
use of these new GAIA data, temperature calibrations will be
refined.
%----------------------Table 4--------------------------
\begin{table}
\begin{center}
\caption{The absolute parameters derived by the non-direct method.}
\label{t4}
\begin{tabular}{ccc}     % 3 columns
\hline \hline
Parameter         &  Primary       &  Secondary         \\
\hline
$M$ (M$_\odot$)   &  $5.0\pm 0.2$  &  $4.05\pm0.1$      \\
$q(M_2/M_1$)      & \multicolumn{2}{c}{$0.81\pm0.01$} \\
$R$ (R$_\odot$)   & $3.1\pm 0.2$   &  $3.8\pm0.1$       \\
log $L$ (L$_\odot$)   & $2.79\pm 0.05$   &  $2.42\pm0.05$  \\
log $g$           & $4.17\pm 0.01$ &  $3.89\pm0.01$     \\
$a$ (R$_\odot$)   & \multicolumn{2}{c}{10.2 $\pm $ 0.4}   \\
d [pc]            & \multicolumn{2}{c}{$2040\pm150$}  \\
\hline\hline
\end{tabular}
\end{center}
\end{table}
%-------------------------------------------------------

\begin{figure}
\centerline{\includegraphics[width=\hsize,clip=]{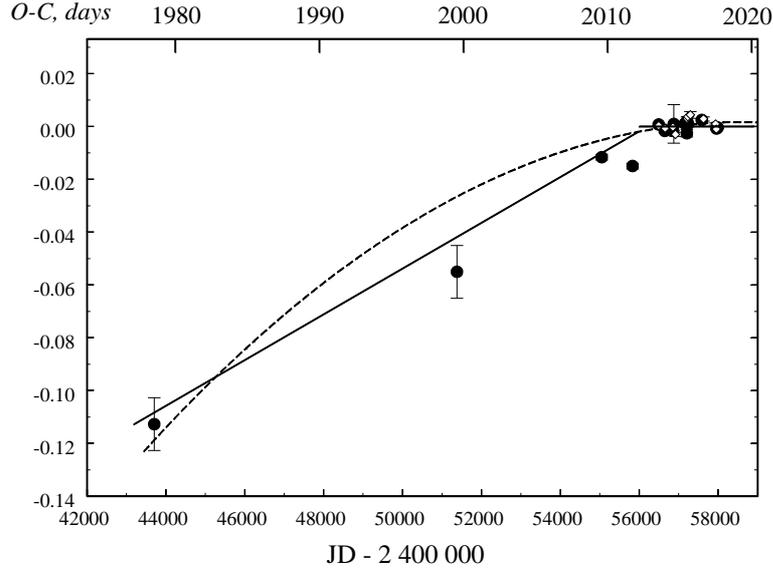}}
\caption{A plot showing the ETV diagram for V2247~Cyg, constructed
using ephemeris (4). The circles represent primary minima, the
diamonds -- secondaries. The data are fitted by a parabolic law
(6) -- the dotted line, and linear laws (4) and (5) -- solid straight
lines.} \label{etv1}
\end{figure}

To derive precise minima times from photoelectric and CCD
observations, we fitted the synthetic LCs, obtained during single
overnight runs, by means of the PHOEBE program varying only the
specific epoch. Sometimes, when only parts of the minimum were
available for the close dates of observations, we compiled them
into one minimum and assigned the acquired minimum time to the
night with more numerous observations. In the case of simultaneous
observations in several filters, the minima times were weighted
and mean values were calculated. The minima times are listed in
Table~\ref{t5}, together with the already published ones. Primary
and secondary minima times were used to construct an ETV diagram,
see Fig.~\ref{etv1}. One can see that the period of the system is
changing: close to JD=2456000, it became shorter.
%
%----------------Figure 6---------------------------------
%----------------------------------------------------------
%

The $O-C$ residuals in the ETV diagram (Fig.~\ref{etv1}) and in
Table~\ref{t5} were calculated using the linear ephemeris which is suitable for the modern epoch:
\begin{center}
HJD ${\rm Min~I~} = 2456857.3783(2)+1.2548450(3)\times E$.~~~~~~~~~(4) \\
\end{center}
The secondary component fills its Roche lobe and the mass
transfer, or mass loss, might be responsible for the period change.
Timings before JD2456000 satisfy the formula:
\begin{center}
HJD ${\rm Min~I~} = 2456857.3808(6)+1.2548564(4)\times E$,~~~~~~~~~(5) \\
\end{center}
which coincides with the KG and Otero (2008) within
errors. The minima times in the ETV diagram can also be fitted by
a parabola (a continuous period change), represented by the
following ephemeris:
\begin{center}
HJD${\rm Min~I~}=2456857.3768(3)+1^d.2548481(9)\times E-8.57(4)\cdot 10^{-10}\times E^2$.~~~~(6) \\
\end{center}
%

%--------------------Figure 7-----------------------
\begin{figure}
\centerline{\includegraphics[width=\hsize,clip=]{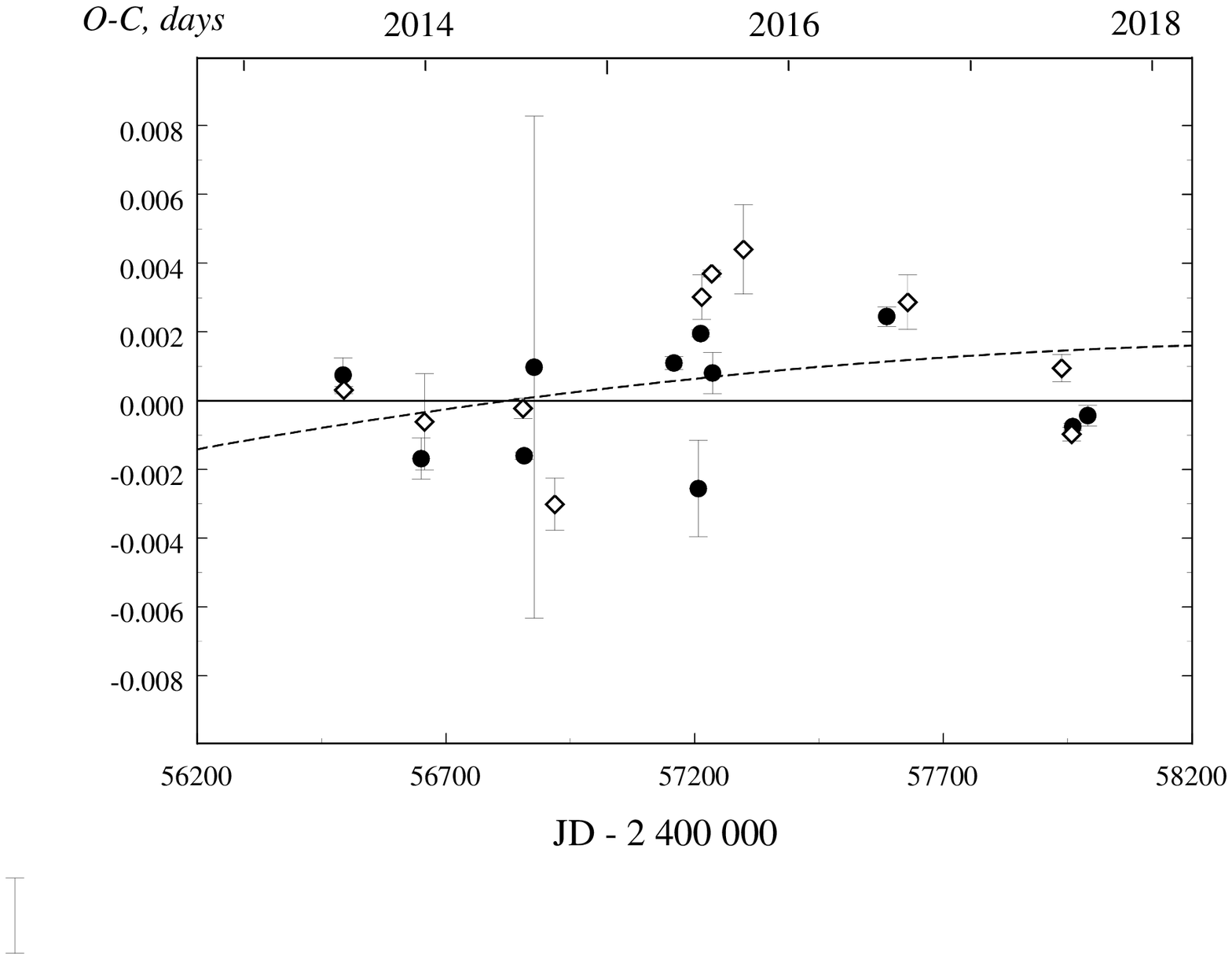}}
\caption{The ETV diagram for V2247~Cyg for the modern epoch on a large
scale. The signs are the same as in Fig.~\ref{etv1}. The parabolic and
linear laws fit observations equally well.} \label{etv2}
\end{figure}
%---------------------------------------------------

%---------------------Table 5-------------------------
\begin{table}
%\tiny
\begin{center}
\caption{Times of minima for V2247~Cyg. The errors for the first two
minima were not available and were set equal to the half of the
least significant digit adduced by the authors.}
\label{t5}
\begin{threeparttable}
\begin{tabular}{lrccc}
\hline \hline
HJD$-$2400000 & Epoch &Residuals&Residuals& Source \\
                &       &from linear& from para-&         \\
 & &ephemeris (4)&bolic fit (6)&               \\
\hline
43706.49       & -10480   & $-$0.1127  & $ $0.0060 & 1 \\
51378.67       &  -4366   & $-$0.0550  & $-$0.0285 & 2 \\
55050.3899(11) &  -1440   & $-$0.0116  & $-$0.0065 & 3 \\
55833.4098(11) &   -816   & $-$0.0150  & $-$0.0126 & 4 \\
56493.4740(5)  &   -290   & $ $0.0008  & $ $0.0014 & t.w., $V$ \\
56495.3549(1)  &   -288.5 & $ $0.0003  & $ $0.0010 &     "     \\
56650.3272(6)  &   -165   & $-$0.0017  & $-$0.0013 & 5 \\
56657.2290(14) &   -159.5 & $-$0.0006  & $-$0.0003 &     "     \\
56855.4949(3)  &     -1.5 & $-$0.0002  & $-$0.0003 & t.w., $UBVRcIc$ \\
56857.3767(1)  &      0   & $-$0.0016  & $-$0.0017 &     "     \\
56877.457(7)   &     16   & $ $0.0010  & $ $0.0009 & 6 \\
56919.4892(8)  &     49.5 & $-$0.0030  & $-$0.0032 & 7 \\
57158.5422(2)  &    240   & $ $0.0011  & $ $0.0005 &     "     \\
57207.4775(14) &    279   & $-$0.0026  & $-$0.0032 & 8 \\
57212.5014(1)  &    283   & $ $0.0020  & $ $0.0013 & t.w., $BVRcIc$ \\
57214.3838(7)  &    284.5 & $ $0.0030  & $ $0.0024 & 7 \\
57234.4620(1)  &    300.5 & $ $0.0037  & $ $0.0030 & t.w., $UBVRcIc$ \\
57236.3423(6)  &    302   & $ $0.0008  & $ $0.0001 & t.w., $U$ \\
57298.4598(13) &    351.5 & $ $0.0044  & $ $0.0036 & 9 \\
57586.4457(3)  &    581   & $ $0.0025  & $ $0.0013 & 7 \\
57628.4825(8)  &    614.5 & $ $0.0029  & $ $0.0017 &     "     \\
57938.4273(4)  &    861.5 & $ $0.0009  & $-$0.0005 & t.w., $UBVRcIc$ \\
57958.5029(2)  &    877.5 & $-$0.0010  & $-$0.0024 &     "     \\
57960.3863(1)  &    879   & $-$0.0008  & $-$0.0022 &     "     \\
57990.5029(3)  &    903   & $-$0.0004  & $-$0.0019 & t.w., $UBV$, PMT \\
\hline\hline
\end{tabular}
Notes. 1 - KG; 2 - Otero (2008); 3 - Hubscher et al. (2010); 4 - Hubscher et al. (2013); 5 - Hubscher (2014); 6 - Hubscher \& Lehmann (2015); 7 - Zasche et al. (2017); 8 - Hubscher (2016); 9 - Hubscher (2017); t.w. - this work.
\end{threeparttable}
\end{center}\end{table}
%
%
%--------------------Figure 8--------------------------------
\begin{figure}
\centerline{\includegraphics[width=\hsize]{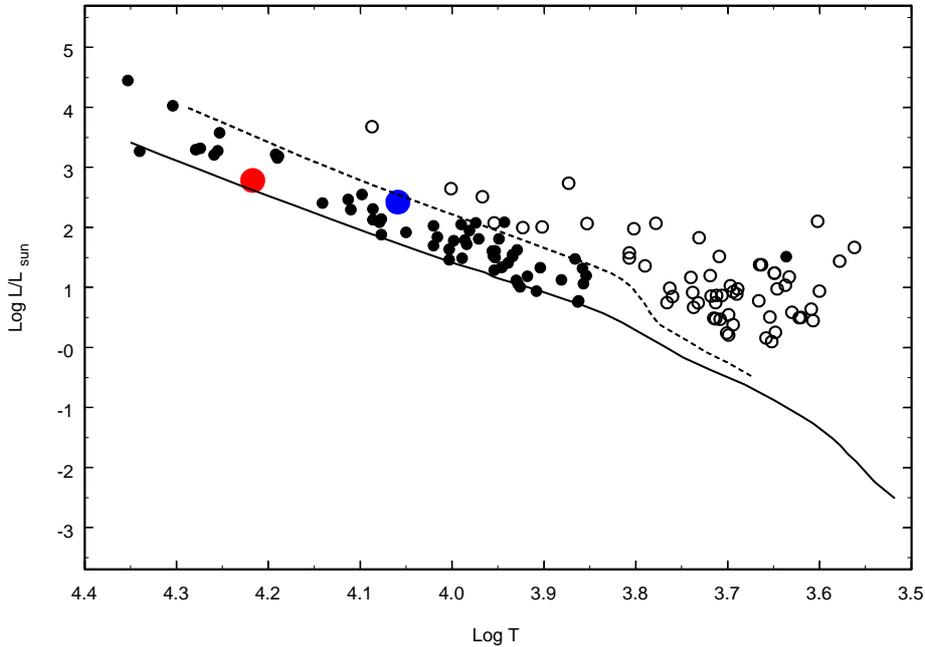}}
\caption{A plot showing the HR diagram for the primary (filled circles) and
secondary (open circles) components of semidetached Algol binaries
taken from Ibano\v{g}lu et al. (2006). Solid and dashed lines
represent the ZAMS and TAMS for solar chemical abundance
(Girardi et al., 2000), respectively. We added the location
of the primary (the red circle) and secondary (the blue circle) components
of V2247~Cyg.} \label{HR}
\end{figure}
%----------------------------------------------------------
%
%--------------------Figure 9-------------------------------
\begin{figure}
\centerline{}
\includegraphics[width=\hsize]{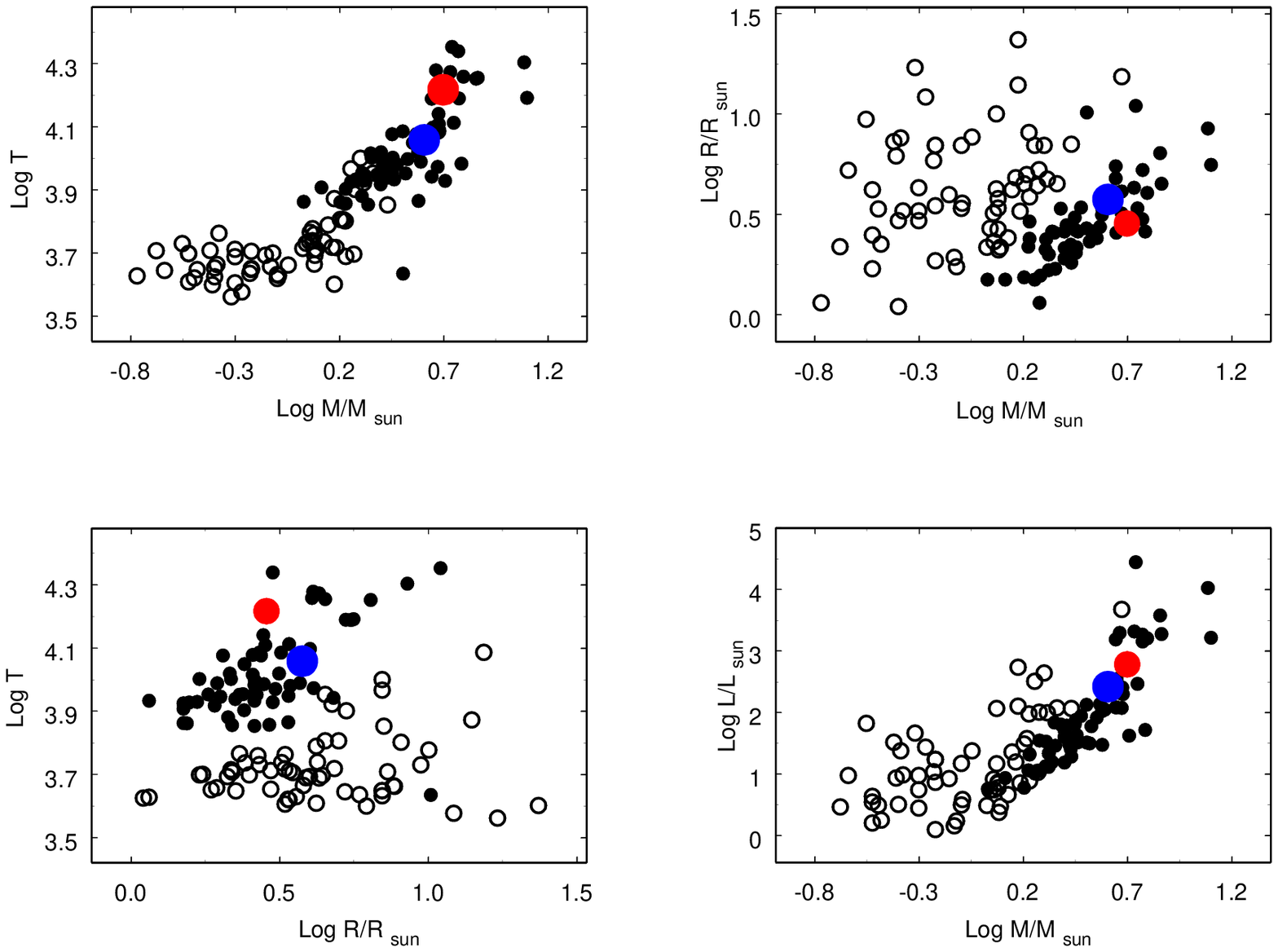}
\caption{Various diagrams for semidetached Algol binaries taken
from Ibano\v{g}lu et al. (2006). Symbols are the same as in
Fig.~\ref{HR}.} \label{diagrams}
\end{figure}
%-------------------------------------------------------------

The weights for the minima times were set equal to $1/\sigma^2$.
Usually for such a configuration of the system one could expect
mass transfer from the secondary component to the primary through
the inner Lagrange point $L_1$. In this case there should be a
gradual increase in the orbital period of the binary star. But we are
watching the opposite case -- the period is decreasing. This means
that a non-conservative mass loss from the secondary through the
$L_2$ Lagrange point takes place. We can't rule out the abrupt
period decrease close to JD~=2456000. Fig.~\ref{etv2} presents the
ETV diagram for the modern epoch on a large scale. We cannot say that
the parabolic fit has any advantage comparing to the linear law.
The linear fit explains old observations better than the parabola
does, see Fig.~\ref{etv1}. So we may suppose that the processes of
mass loss and mass transfer in this system can occur episodically but not continuously. Some systematic difference is seen
in the shape of the $V$ light curve observed in 2013 comparing to LCs
obtained in the 2014-2017 interval, i.e., the primary minimum was
deeper by 0.01 in 2013. This difference may also be assigned to
the processes mentioned above.

Figs.~\ref{HR} and~\ref{diagrams} show the location of the primary
and secondary components in the HR diagram and $\log T-\log M$,
$\log R-\log M$, $\log T-\log R$, and $\log L-\log M$ diagrams for
V2247~Cyg as well as for 61 semidetached Algol binaries with
well-determined absolute parameters (Ibano\v{g}lu et al., 2006).
While the position of the primary component is similar to that of
other primaries, the secondary of V2247~Cyg lies near the border,
of or even outside, the region occupied by other secondaries. It's
worth mentioning that, among 61 semidetached binaries listed in
Ibano\v{g}lu et al. (2006), none of the systems consists of two
B-class stars. What makes V2247~Cyg unusual, when compared to other
semidetached Algol-type binaries, is its higher mass and
temperature of the secondary.

\section{Conclusions}

Using multicolour photometry we obtained reliable parameters for
the Algol-type binary V2247~Cyg: colour indices, interstellar
reddening, mass ratio, inclination, effective temperatures of the
components, fractional radius, the potential, as well as the
albedo of the secondary and the gravity brightening coefficients.
V2247~Cyg was found to be a semidetached system with the secondary
filling its Roche lobe.

The B5III-V spectral type was ascribed to the low-resolution
spectrum of V2247~Cyg.

Due to the lack of data on radial velocities, the masses of the
components were computed by a non-direct method. The mass ratio
derived this way is in excellent agreement with the value obtained
through solving LCs.

The study of the ETV diagram enabled us to discover the orbital period
decrease. It can be explained by a mass loss from the less massive
secondary component.

In the HR and $\log T-\log M$, $\log R-\log M$, $\log T-\log R$,
and $\log L-\log M$ diagrams the primary component of V2247~Cyg
lies well within the region occupied by the primaries of
semidetached Algol-type binaries, whereas its secondary
differs from other secondaries due to its higher mass and
temperature.

We'd like to encourage high-resolution and high signal-to-noise
spectroscopic observations of the system in order to determine the
masses of the components from radial velocity curves.

\begin{acknowledgements}
This study was partly supported by the scholarship of the Slovak
Academic Information Agency(IMV), RNF grant 14-12-00146 and RFBR
grant 18-502-12025(IMV). We are grateful to professor N.~N.~Samus
for providing access to the card catalogue of SAI, Dr.
V.P.~Goranskij for permission to use his data prior to publication
and to an anonymous referee for important corrections.
\end{acknowledgements}

% for the bibliography, at the end

%\bibliographystyle{caosp} % style aa.bst
%\bibliography{V2247} % your references Yourfile.bib
%

\end{document}